# Mapping Conductance and Switching Behavior of Graphene Devices In Situ

*Ondrej Dyck,1\* Jacob L. Swett,2\* Charalambos Evangeli,2 Andrew R. Lupini,1 Jan A. Mol,2,3 Stephen Jesse1*

1 Center for Nanophase Materials Sciences, Oak Ridge National Laboratory, Oak Ridge, TN
2 Department of Materials, University of Oxford, Oxford OX1 3PH, UK
3 School of Physics and Astronomy, Queen Mary University of London, London E1 4NS, UK

*\* These authors contributed equally*

**Abstract**
Graphene has been proposed for use in various nanodevice designs, many of which harness emergent quantum properties for device functionality. However, visualization, measurement, and manipulation become non-trivial at nanometer and atomic scales, representing a significant challenge for device fabrication, characterization, and optimization at length scales where quantum effects emerge. Here, we present proof of principle results at the crossroads between 2D nanoelectronic devices, e-beam-induced modulation, and imaging with secondary electron e-beam induced currents (SEEBIC). We introduce a device platform compatible with scanning transmission electron microscopy investigations. We then show how the SEEBIC imaging technique can be used to visualize conductance and connectivity in single layer graphene nanodevices, even while supported on a thicker substrate (conditions under which conventional imaging fails). Finally, we show that the SEEBIC imaging technique can detect subtle differences in charge transport through time in non-ohmic graphene nanoconstrictions indicating the potential to reveal dynamic electronic processes.

**Introduction**
Scanning transmission electron microscopy (STEM), through the advent of effective aberration correction, is now able to probe the position, bonding, and elemental composition of single atoms in materials.[1] These capabilities have been complemented by the introduction of standard tools and techniques for *in situ* experimentation, which have both advanced in capability and ease over the past decade.[2] The combination of these capabilities has made the STEM a powerful tool to understand nanoscale electronics and materials broadly, among other applications. Additionally, there has been an increasing push to develop the STEM as an atomic-scale fabrication tool, shifting the instrument from solely a characterization tool to a fabrication tool, analogous to the differences between a scanning electron microscope (SEM) and an electron beam lithography (EBL) system.[3,4] The aim of these fabrication efforts is to structure matter first at the nanoscale and then the atomic scale, with the goal of developing new electronic, quantum, and nanoelectronic devices.[5–12] However, in nanoscale electronics and molecular electronics, the flow of charge carriers is of primary interest for the function of the devices. While *in situ* biasing techniques have allowed this aspect to be probed in tandem to characterization via more common techniques, it does not permit spatial understanding of the conductivity and only permits a rudimentary understanding of connectivity (e.g. whether the device is connected to both leads or not). One mechanism to perform such direct spatial characterization and mapping of connections is electron beam included current (EBIC) imaging. We will focus, in particular, on a sub-class of EBIC known as secondary electron EBIC (SEEBIC) imaging, which is the induced generation of a current via emission of secondary electrons from an impinging electron beam[13] and is sensitive enough to detect single layers of graphene.[14] This work builds on both the work in advancing STEM secondary electron (SE) imaging[15] and STEM SEEBIC. We will begin with a description of the operando platform, describe the basic physics behind SEEBIC image formation and then examine several examples illustrating the usefulness of the method for characterization of graphene-based nanodevices.

**Experimental Description**

Figure 1 (a) shows renderings of an overview of the device design. The inset on the left shows the full chip. The main background image shows the central region that has an electron transparent, 20 nm thick, $SiN_x$ membrane support over which the devices are fabricated. The inset image on the right shows a magnified view of a graphene nanoribbon device suspended over an aperture in the $SiN_x$ membrane. Contact pads facilitate electrical connection to a Protochips™ electrical holder to facilitate operando experimentation in the STEM. The base platform can be adapted to a variety of different experimental designs. A detailed description of the general fabrication procedure can be found in the supplemental information as well as a description of the specific device design used here. Transfer of graphene grown on Cu foil via chemical vapor deposition was accomplished on the wafer scale by Graphenea (San Sebastián, Spain).

While suspended regions of the graphene are amenable to high resolution imaging, graphene in supported regions of the device and in devices without apertures is almost completely invisible to conventional medium/high angle annular dark field (M/HAADF) imaging by virtue of the fact that the graphene signal is mixed with the signal from the 20 nm silicon nitride window. This situation can be resolved through the use of SEEBIC imaging as will be described. Next we will introduce the SEEBIC imaging technique and distinguish it from other similar variants like the more common EBIC technique.

EBIC imaging is a technique that has been used widely in semiconductor fabrication (particularly for failure analysis) and has been employed as a characterization method for over fifty years.[16] Broadly speaking, EBIC techniques attempt to spatially measure currents occurring within a sample driven by a focused primary e-beam in a scanning electron microscope (SEM). The most well-know EBIC technique, shown in Figure 1 (b), uses an electron beam to measure electron-hole (e-h) pair recombination lengths and maps the depletion region in a semiconductor p-n junction. Because each incident electron can generate multiple e-h pairs the resultant signal current, $I_{EBIC}$, can exceed the primary beam current, $I_{Beam}$. A variation of this technique, shown in Figure 1 (c), measures absorbed electron current from the beam and can be used in probing disconnects in a non-destructive manner, when accelerating voltage is kept low. This configuration we term electron beam absorbed current (EBAC) imaging. In this configuration the signal current generated, $I_{EBAC}$, is less than (or at most equal to) the primary beam current. While both signals are electron beam induced (in the most general sense), the mechanism generating image contrast is different for the two configurations.

For thin samples and with higher primary beam energies, as is the case in a STEM, the beam removes electrons from the sample resulting in a hole current into the amplifier. This is due to e-beam-induced emission of secondary electrons (SE) so we refer to this imaging mode as secondary electron EBIC (SEEBIC), summarized in Figure 1 (c). This follows the naming conventions proposed by Hubbard et al.[13] In this imaging mode, the signal intensity, $I_{SEEBIC}$, is governed by the secondary electron yield of the specimen, the sample interaction volume, and primary beam energy, among other parameters. Because the interaction volume is small for thin samples, and the secondary electron yield is low for beam energies typically used in STEM, $I_{SEEBIC}$ can generally be assumed to be much less than the primary beam current.[17] It should be noted that each of these configurations is dependent upon beam/specimen interactions, electrical connections within the sample and electrical connections to the sample. For example, EBAC is only possible when the beam (or some measurable fraction of the electrons) can be stopped in the conductive portion of the specimen and as such is typically not possible at all in high accelerating voltage STEM imaging where almost all electrons, as the name suggests, transmit through the sample. Likewise, SEEBIC depends on electrons being emitted from the surface, which depends on the escape depth for secondary electrons for that material. The probability for emission decays exponentially with depth and is therefore primarily confined to emission from the top (or bottom) several nanometers of the specimen.[17] Because of the limited interaction volume, SEEBIC is often higher resolution, but typically at the expense of less signal current.

Detecting the SEEBIC signal depends on a conductive pathway to the amplifier. In the configuration shown in Figure 1 (d), the entire conductor would appear of similar intensity. Figure 1 (e) shows a conceptual diagram of the SEEBIC signal intensity that would be observed when imaging two conductors with a resistive element connecting them. The conductor connected to the amplifier appears bright (higher signal), while the conductor connected to ground appears dark (lower signal). Technically speaking, both conductors are grounded but we are measuring the current flow on the amplifier side and not the other side, which we will refer to as the grounded side. SEs are generated by the primary beam in both regions equally, however the grounded electrode can dissipate charge accumulation through the path to ground which is not detected by the amplifier. In the resistor/insulator region charge flow is hampered by the material. In this instance the relative resistivity in the pathway to ground determines the SEEBIC intensity. Reversing the connections produces a complementary inverse image. This variant of SEEBIC imaging is akin to resistive contrast imaging but with the current flowing in the opposite direction.[18,19,13]

To acquire SEEBIC images of our devices we connected the electrodes to a Femto DLPCA-200 transimpedance amplifier (TIA) with a gain set to $10^{11}$ V/A and 3 dB bandwidth of 1.1 kHz. The signal from the TIA was mapped according to e-beam position to generate a new image channel that can be acquired in parallel with conventional HAADF imaging. Figure 2 shows a summary of a typical dataset. Figure 2 (a) shows a labeled HAADF image of the device. Note that the supported regions of the graphene (i.e. the graphene directly on the $SiN_x$) are not distinguishable. The IV trace, inset, shows an ohmic response and indicates the device is functioning. Figure 2 (b) shows a labeled SEEBIC image which was acquired concurrently with the HAADF image. In this image we can now clearly observe the electrically connected and conductive regions of the device. Two example profiles were extracted from the SEEBIC data and are displayed in Figure 2 (c). A significant amount of spurious interference was present in all the SEEBIC images collected. This interference was found to be synced to the electrical mains and was removed as described in the supplemental materials.

Several features are worthy of comment. In the suspended region, at the center of the profiles, we have a reference intensity for vacuum and can clearly distinguish this from the intensity of the suspended graphene. While we refer to this as "graphene" it should be noted that this signal is likely significantly enhanced by surface contaminants that are nearly always present on graphene.[14] The signal from bare $SiN_x$ is not substantially different from vacuum. As $SiN_x$ is an insulator, a positive charge will build up quickly and act to recapture SEs thus exhibiting no net signal into the TIA. The graphene on $SiN_x$ region shows a pronounced increase in signal above the suspended graphene region. This is because SEs emitted from both the graphene and $SiN_x$ can be replenished through the electrically conductive path to the TIA. The signal is again increased at the electrical contacts, however, perhaps less than one might expect especially when compared to the increase in intensity observed in the HAADF image. The contrast in HAADF is generated by total scattering of the primary beam. But the contrast in SEEBIC is generated by emission of secondary electrons, mostly near the surface. Thus the sample thickness, elemental mass, and work function of the material in this region plays a less significant role in contrast generation in SEEBIC compared to HAADF.

**Resistive Contrast Imaging**

SEEBIC is only possible with a conductive path from the region being imaged to the TIA. This gives the technique substantial utility in probing electrical conductance and connectivity and can be useful in diagnosing aspects of the device that could not otherwise be observed through more common STEM imaging mechanisms (e.g. HAADF/MAADF imaging or EELS). Here, we discuss how SEEBIC can aid understanding of device structure during *in situ* experimentation by providing insight into graphene cracks. *In situ* electrical characterization of the devices shown in Figure 3 permitted understanding of a general failure of the device (open circuit) as shown by the IV traces in Figure 3 (d) and (h). Without the ability to spatially resolve or correlate electrical data to the resolution of the STEM, we are given little insight into the location or nature of the failure. The HAADF image in Figure 3 (a) gives us no additional insight into the failure. However, in this instance the discontinuity in the circuit and the electrical contacts on each side of the device, permit independent SEEBIC imaging of each half, Figure 3 (b) and (c), while keeping the other half grounded in a resistive contrast imaging mode. Such imaging produces a strong SEEBIC signal up to the discontinuity, with a sharp drop in intensity as the beam crosses to the grounded other half of the device. Superimposing the two images gives an intuitive high-resolution map of the device failure, Figure 3 (d). Here, we see that the discontinuity exists at the edge of the bottom contact, an extremely difficult location to detect with other means since the contact itself is also discontinuous at the same location. A second example of this type of characterization is shown in Figure 3 (e)-(h) where a fully supported graphene nanoribbon is examined and found to have a discontinuity on the $SiN_x$ substrate.

**Dynamic Conductance Switching**

These examples illustrate the usefulness of the SEEBIC technique in revealing conductivity and electrical connectivity in static nanodevices. In this case the device properties are constant through time and do not depend on the beam position. However, within the field of nanodevices more generally we are very interested in properties that can be influenced by local changes like electrical or structural variations. One such device configuration is the molecular tunnel junction formed within a graphene nanogap.[20–24] In this configuration a nano-sized gap between two graphene contacts allows an interface with a molecule facilitating a tunneling current through the molecule.[20,22] A fairly reliable method has been established to create such nanogaps through the use of feedback-controlled electroburning.[21,22] Current is passed through a graphene nanoconstriction, like that show in Figure 3 (e)-(h), until joule heating in the narrowest region starts to sublimate and a resistance increase is observed. The voltage is then rapidly ramped down to prevent run-away device failure. This process is repeated until a nanogap is formed.

In some experiments molecules are introduced into the nanogap to create the tunnel junction.[20,22] However, within these graphene nanogaps a reversible conductance switching behavior has been observed in the absence of the introduction of additional molecular species to the junction.[20,25] It was suggested that the formation and destruction of carbon filaments bridging the gap act to facilitate the transport.[20,24] In addition, electroburning has also been observed to spontaneously form graphene quantum dots within the gap acting as single electron transistors.[23,26] A recent study also employed the use of the e-beam to charge an $MoS_2$ device substrate creating a gating field, dependent on beam position.[27] In these examples the device can dynamically change in response to charge accumulation and dissipation.

Since the SEEBIC imaging technique is well positioned to examine conductance, connectivity and charge flow, we performed *in situ* electrical breakdown to create a non-ohmic nanostructure within a graphene nanoconstriction to investigate the SEEBIC response to a dynamically varying system. Figure 4 (a) shows a concurrently acquired HAADF/SEEBIC image pair with an accompanying IV trace acquired prior to the electrical breakdown procedure. The plot in Figure 4 (b) shows the resistance measured during the electrical breakdown procedure where the dramatic increase in resistance over the last few cycles indicates the formation of a non-ohmic nanostructure in the graphene ribbon (this could be a nanogap, quantum dot, series of quantum dots, or various combinations). Figure 4 (c) shows a HAADF/SEEBIC image pair acquired after the electrical breakdown procedure with an accompanying IV trace showing the non-ohmic response. A higher magnification HAADF/SEEBIC image pair of the electrical breakdown region are shown in Figure 4 (d). We can clearly observe mass loss from the $SiN_x$ substrate in the HAADF image. The SEEBIC images in Figure 4 (a), (c), and (d) were acquired with both electrodes connected to the TIA. In Figure 4 (e) we connected each side independently to the TIA in resistive contrast imaging mode. In this configuration we no longer observe static differences in conductance as observed previously (Figure 3). Instead we observe strong dynamic signals due to a combination of electron beam gating,[27] substrate charging (also induced by the beam), and voltage induced switching (from varying the potential on the leads and possibly field-induced restructuring of bridging molecules within the gap).[20,23]

In Figure 4 (f) we examine a portion of the image in Figure 4 (e) more closely. The SEEBIC intensity is plotted as a function of time, instead of beam position, and we can clearly observe the existence of telegraph noise. A Gaussian mixture model clustering routine was used to cluster the intensities and a kernel density estimate of the clusters is shown on the right side of the plot illustrating the degree of separation in intensity between the two states. This indicates a reversible and sustained conductance switching is occurring during the imaging process. The companion image shown in the lower portion of Figure 4 (e) displays a similar switching behavior over various portions as well as variations on a longer timescale and many more switching states (i.e. SEEBIC intensities). Intensities in this image are more closely examined in the supplemental information.

The specifics of the beam induced switching require and deserve additional study. What we stress here is that these dynamics can be *detected* using SEEBIC. However, we can postulate several mechanisms that could produce such behavior. First, as is clear from SEEBIC imaging, the beam interacting with the sample generates an excess of charge that is replenished when the device is connected to ground. Given the high resistivity of graphene devices, this will result in a voltage relative to ground, which, if excessive, will result in modification of the device. Second, electron beam gating was recently demonstrated by Das and Drndić.[27] In this work they show how e-beam induced substrate charging can act as a gate that can be switched off and on through e-beam irradiation. These results begin to show how the electron beam could readily modulate the transport by merely being in the vicinity of a charge sensitive (i.e. field effect sensitive) device. Taken together, we see that by inducing a voltage and then modulating the conductance, dramatic changes in transport can be observed and spatially correlated to the beam position to map the conductance of the specimen.

**Conclusion**
These collective results point toward future directions for STEM SEEBIC to analyze conductance with spatial resolution *in situ*. The results also have important implications for routine STEM characterization of nanoelectronic devices, since the transport measured here suggests that device modification beyond beam damage and other known effects[28] may occur during imaging solely due to the buildup and movement of charge. However, as the nature of these results hint, further understanding of the mechanisms and improved protocols will be necessary to probe the richness of device transport. These results hopefully help to steer that research on a productive path.

**Acknowledgment**

This work was supported by the U.S. Department of Energy, Office of Science, Basic Energy Sciences, Materials Sciences and Engineering Division (O.D. A.R.L., S.J.), and was performed at the Center for Nanophase Materials Sciences (CNMS), a U.S. Department of Energy, Office of Science User Facility (O.D.). J.A.M. was supported through the UKRI Future Leaders Fellowship, Grant No. MR/S032541/1, with in-kind support from the Royal Academy of Engineering. The authors acknowledge use of characterization facilities within the David Cockayne Centre for Electron Microscopy, Department of Materials, University of Oxford, alongside financial support provided by the Henry Royce Institute (Grant ref EP/R010145/1).

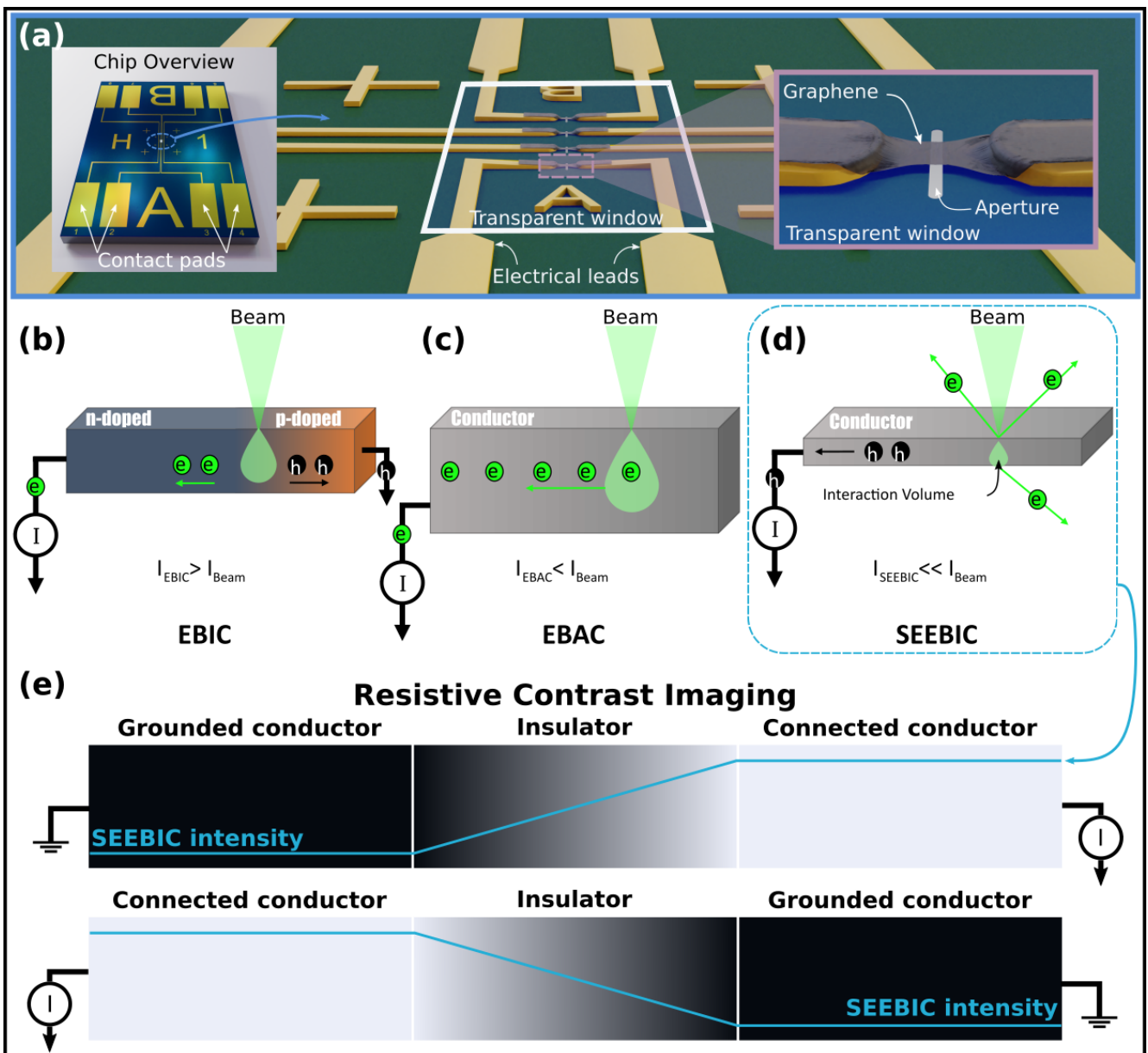

Figure 1: Operando STEM device overview and introduction to the SEEBIC imaging modality. (a) Renderings of the device platform at various magnifications. (b)-(d) Schematic drawings comparing the charge flow involved in contrast generation for EBIC, EBAC, and SEEBIC respectively. (e) Schematic drawings illustrating the SEEBIC intensity one would expect to observe across a resistive element between two conductors, one grounded and one connected to a transimpedance amplifier.

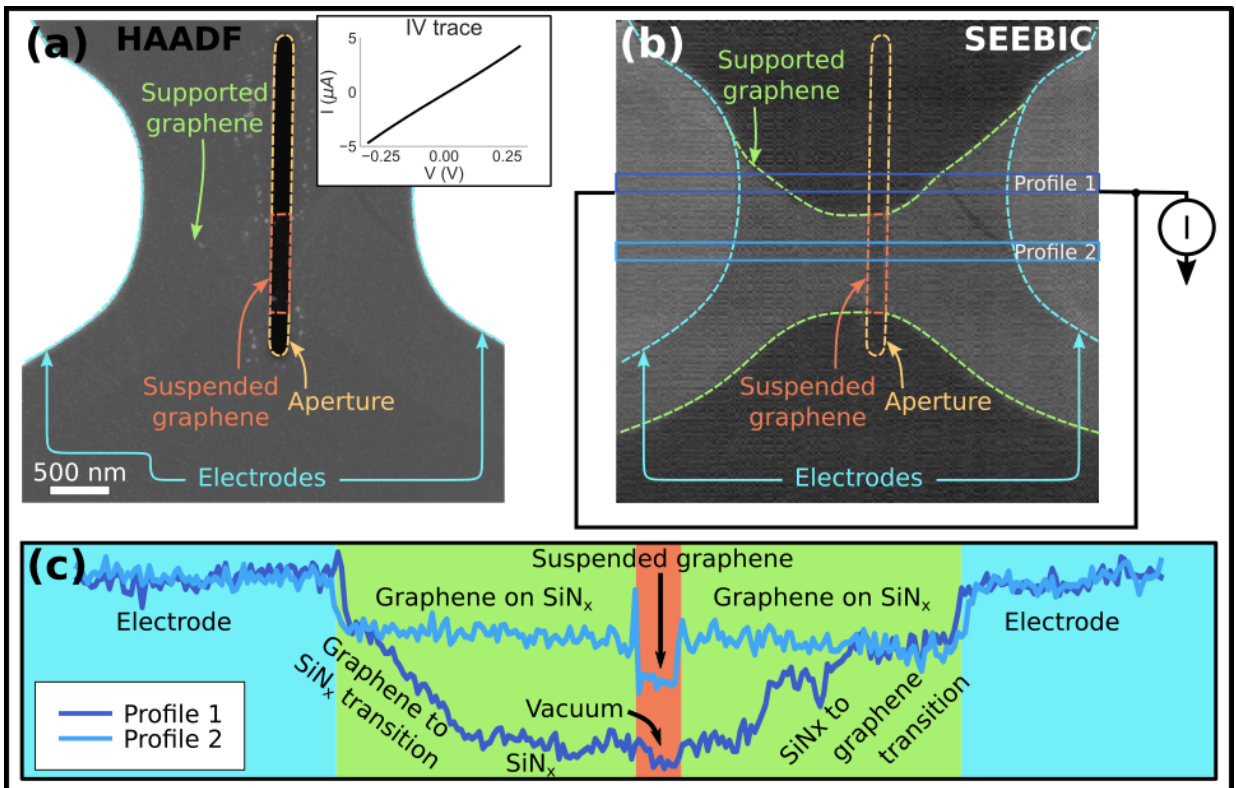

Figure 2: Example HAADF/SEEBIC image dataset. (a) HAADF image of a graphene device with various features labeled. Inset on the upper right is the IV trace acquired during examination. (b) SEEBIC image acquired simultaneously with the HAADF image. The supported graphene is now clearly visible (outlined in green). (c) SEEBIC intensity profiles from the locations marked in (b).

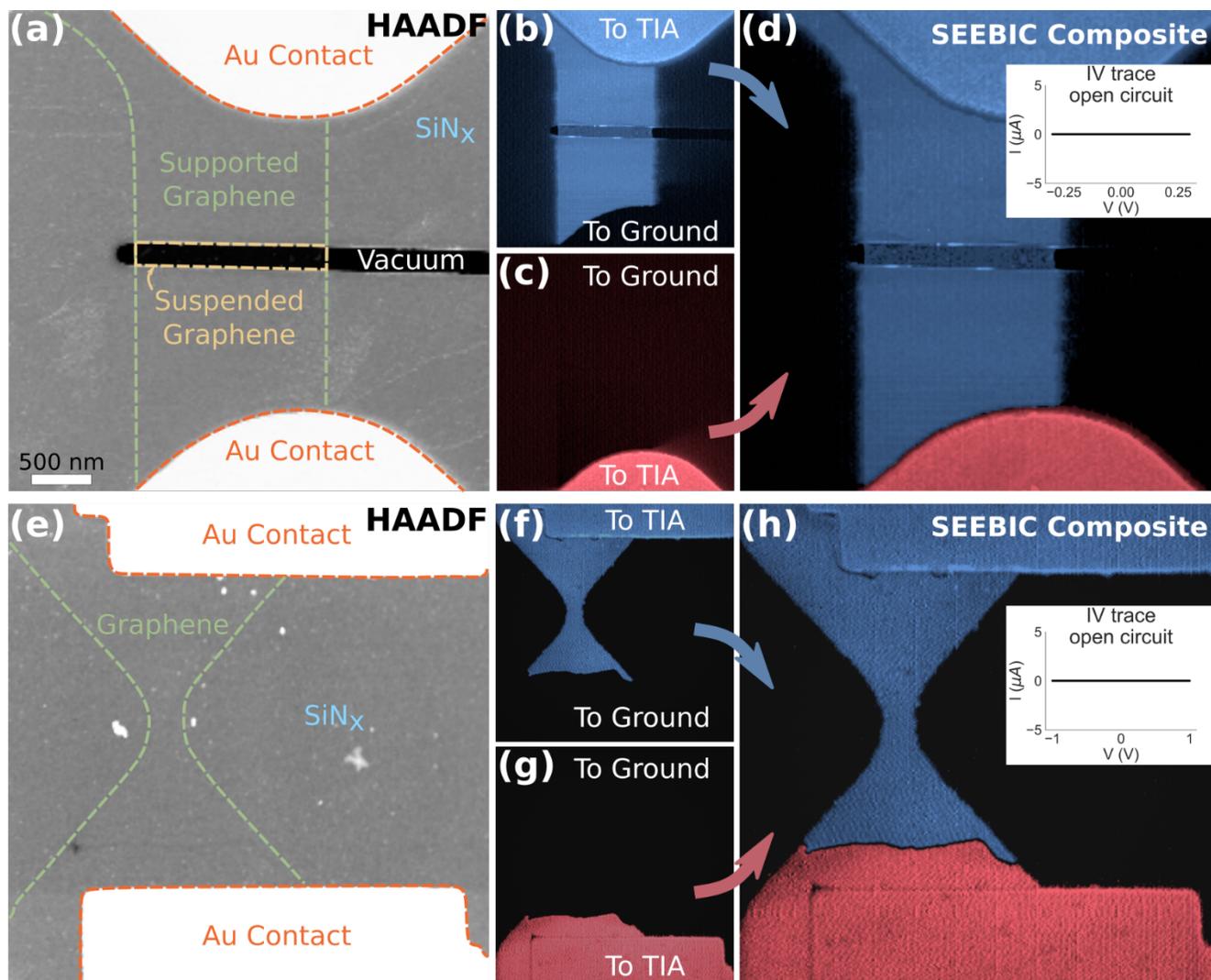

Figure 3: Example of SEEBIC open circuit device diagnostics. (a) HAADF-STEM image of a suspended graphene device with labeled features. (b) and (c) artificially colored SEEBIC images acquired with the transimpedance amplifier connected to alternate electrodes as indicated. (d) Composite SEEBIC image where color denotes the parent image, (b) and (c). (e)-(f) second example on a fully supported graphene nanoribbon.

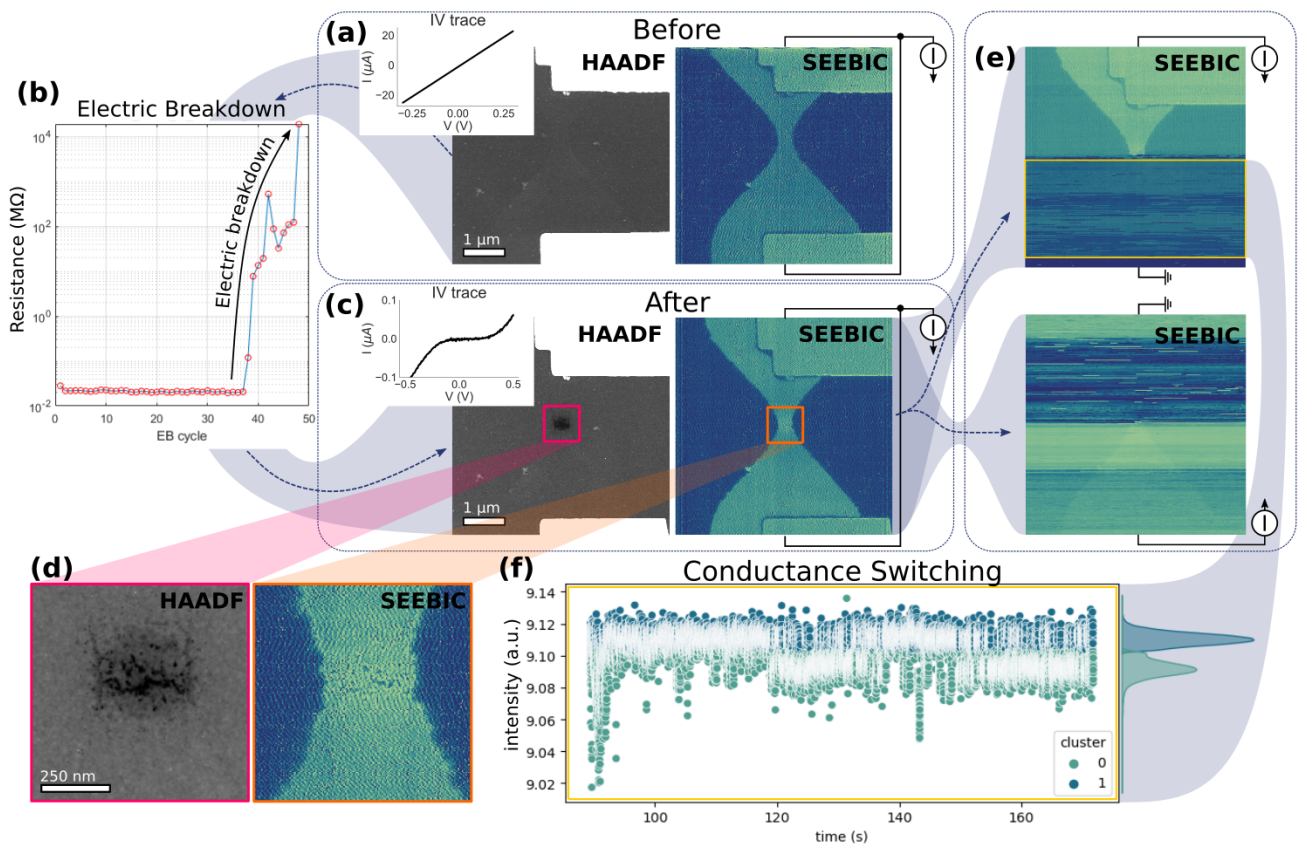

Figure 4: Observation of conductance switching in a graphene nanogap using SEEBIC. (a) Initial sample state. A concurrently acquired HAADF/SEEBIC dataset shows the electrode and graphene nanoribbon location. The electrical connections to the TIA for the SEEBIC acquisition are shown schematically around the SEEBIC image. The IV trace shows the initial ohmic response of the device. (b) Plot of the device resistance recorded during the electric breakdown process to form a non-ohmic nanostructure. (c) Sample state after the electric breakdown process. The IV trace now illustrates the non-ohmic device response. (d) magnified views of the region where the non-ohmic nanostructure has formed. (e) SEEBIC images acquired by connecting each electrode individually to the TIA and the opposite electrode to ground. (f) plot of the SEEBIC intensity observed as a function of time illustrating a reversible change in conductance. The intensities were clustered using a Gaussian mixture model and the kernel density estimates of each cluster are shown plotted on the right.

# Mapping Conduction and Switching Behavior of Graphene Devices In Situ

*Ondrej Dyck,[1]\* Jacob L. Swett,[2]\* Charalambos Evangeli,[2] Andrew R. Lupini,[1] Jan A. Mol,[2,3] Stephen Jesse[1]*

[1] Center for Nanophase Materials Sciences, Oak Ridge National Laboratory, Oak Ridge, TN
[2] Department of Materials, University of Oxford, Oxford OX1 3PH, UK
[3] School of Physics and Astronomy, Queen Mary University of London, London E1 4NS, UK

*\* These authors contributed equally*

*Keywords: Secondary electron e-beam induced current, graphene, scanning transmission electron microscopy, secondary electron yield*

## Supplemental Information

### General fabrication of suspended silicon nitride devices

The most basic suspended silicon nitride device consists of $SiN_x$ deposited on a Si wafer anisotropically etched from the backside as shown in Figure S1. The simplest fabrication method employs <100> Si, which permits the anisotropic etching. To define a suspended window on one side of the device, the $SiN_x$ is patterned via conventional photolithography, then removed using reactive ion etching (RIE). This serves to create a hardmask for subsequent processing. Given the nature of the anisotropic etch, one must simply take into account the thickness, T, of the wafer and the angle of the etch, which is 54.74°, and then define the correct initial window width, $W_i$, on the opposite side to achieve the desired final window width, $W_f$.

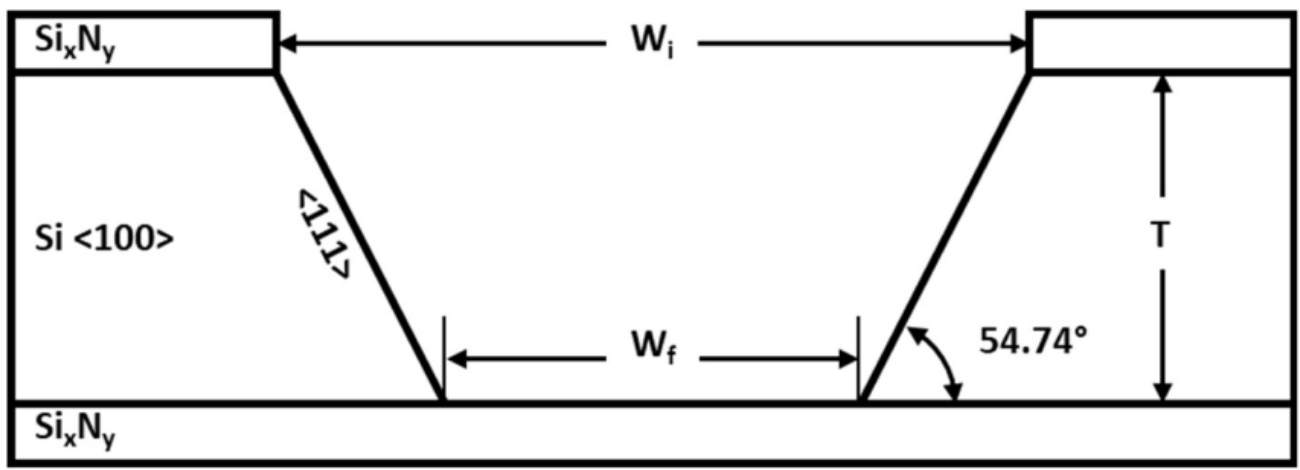

Figure S1: Schematic showing anisotropic etching of <100> Si. Using a $SiN_x$ hard mask and knowing the wafer thickness T, the desired window width $W_f$ can be defined from the hard mask open $W_i$.

Etching is commonly done with KOH (although TMAH may also be used), with a standard recipe using a 30% KOH/water solution at 80° C, for which etching of a 300 µm wafer will take approximately four hours.[1] To protect features possibly fabricated on the membrane side of the wafer, it is possible to load wafers into a single-sided wafer holder, as shown in Figure S2. In this setup, wafers are sealed with O-rings into a compatible holder and only one side of the wafer is exposed to the solution. Combined with a condensing system, wafers can reliably be etched several times in the same solution, producing consistent results.

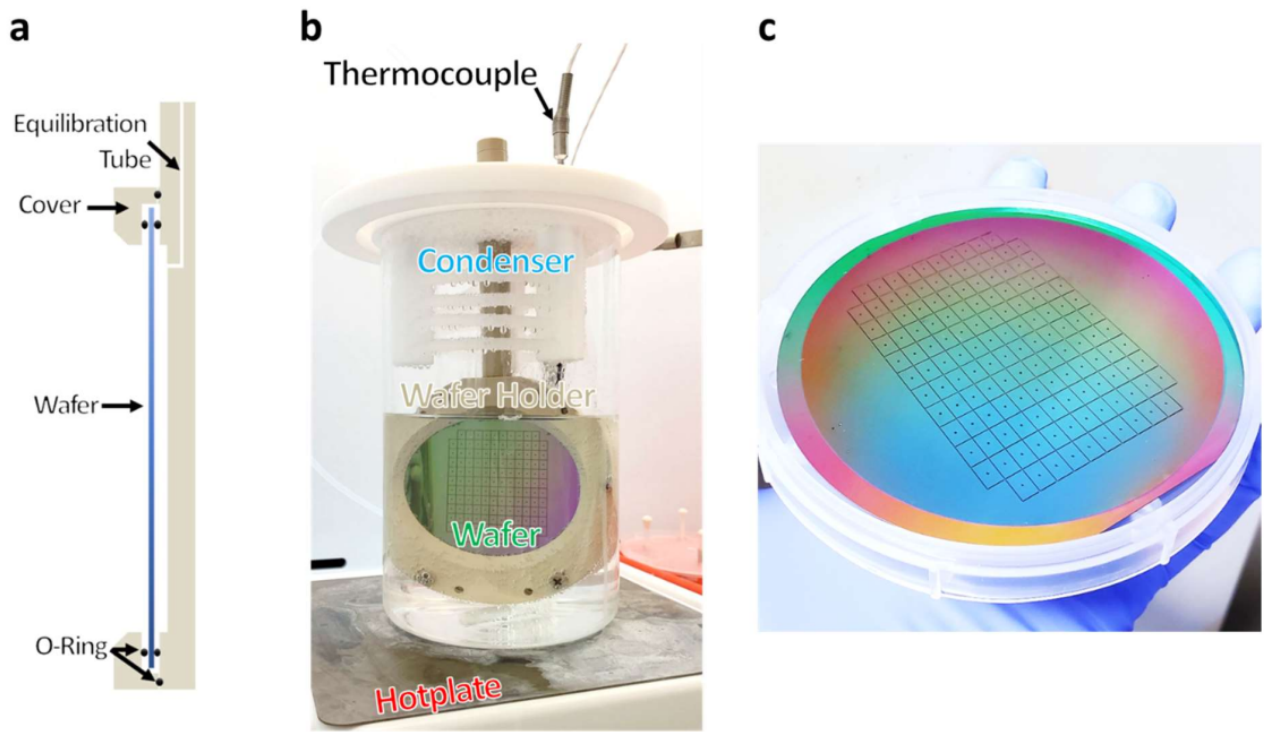

Figure S2: Backside KOH etching. (a) schematic of single-sided wafer holder with backside protection for KOH etching, (b) photograph of wafer being etched on a hotplate with condenser system attached, (c) backside of a wafer after etching.

In addition to windows, other features may be defined with backside KOH etching, such as partial etches through the wafer, termed v-grooves, which help to facilitate separation of devices later. To produce devices, <100> double side polished Si wafers were commercially procured. If low-noise applications are desired, then ideally high-resistivity wafers should be used. Next, a wet thermal oxide layer is grown, with the exact thickness being application dependent. Oftentimes 200 nm - 1000 nm is used. Finally, low-pressure chemical vapor deposition (LPCVD) was used to deposit layers of Si-rich silicon nitride on both sides of the wafer. Low-stress Si-rich silicon nitride was used given its ideal mechanical properties, which lend stability.[2] The above process describes the most basic suspended silicon nitride devices, however there are many permutations possible beyond this. The most common is the inclusion of metal electrodes for biasing and field generation. These electrodes, which can be fabricated via photolithography or electron beam lithography (EBL) or sometimes both, can lead to versatile devices capable of operation of a variety of environments ranging from *in situ* TEM to fluidic cells.[3]

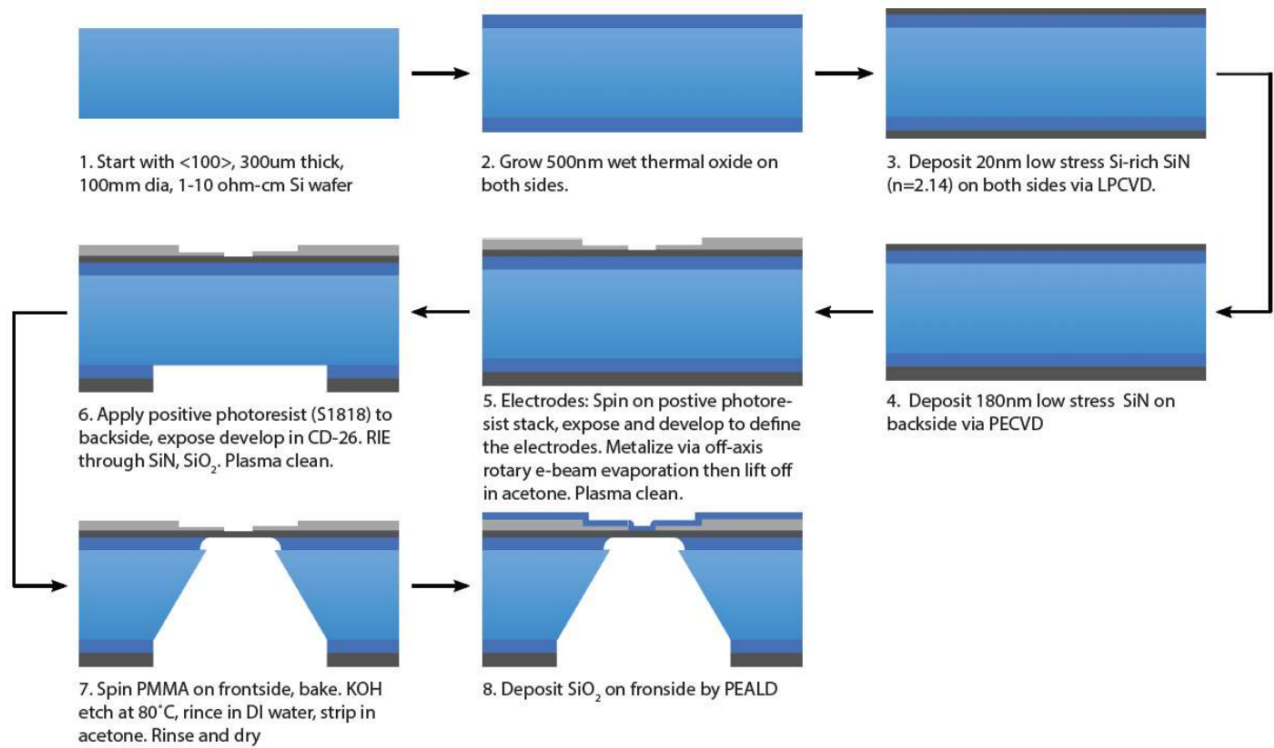

Figure S3: Example fabrication protocol for suspended devices. The final step is optional, based on whether an insulating capping layer is desired. In the experiments shown here, this step was not performed to allow for transfer and patterning of graphene onto the electrodes.

The end result of the fabrication, regardless of the device and crucially, when successful, are beautiful wafers. Some examples are shown in Figure S4.

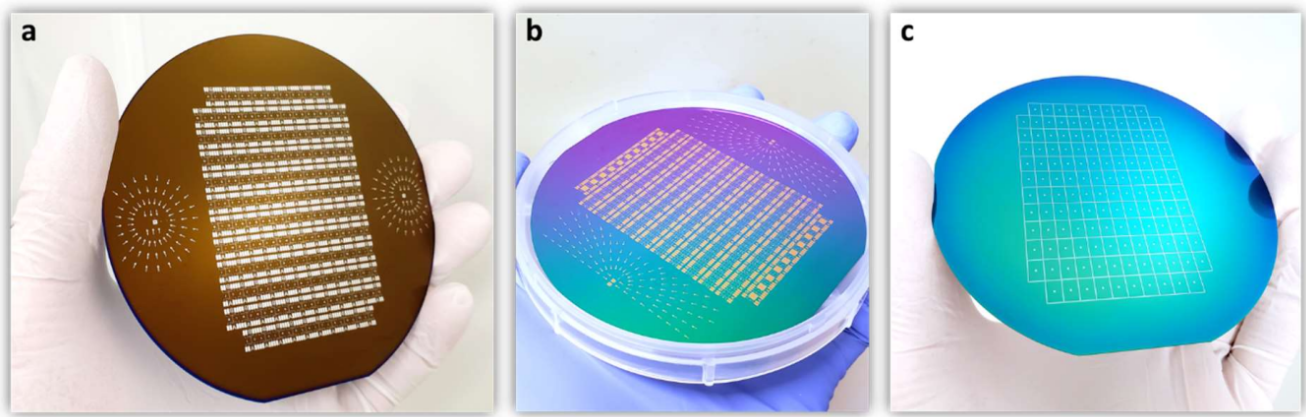

Figure S4: Examples of 100 mm wafers of fabricated suspended $SiN_x$ devices (a) two-terminal electrodes on 50 nm $SiN_x$, (b) dielectrophoresis devices fabricated on 20 nm $SiN_x$ on 1000 nm of $SiO_2$, (c) backside of a suspended photonic device.

**Specific fabrication of silicon nitride devices**

The devices used in this study were fabricated with a thicker underlying dielectric layer of 1000 nm wet thermal oxide to decrease capacitive coupling between the electrodes and the conductive Si substrate. As with other devices, the $SiN_x$ thickness was typically 20 nm. Electrodes were fabricated in two steps. Fine electrodes were patterned via electron beam lithography (EBL) and subsequently metalized via e-beam evaporation (Cr 5 nm/Au 35 nm). Larger electrodes to facilitate contact in the STEM were then patterned via photolithography and again metalized via e-beam evaporation (Cr 5 nm/Au 95 nm). Devices were then patterned from the backside using reactive ion etching to define areas on the backside for subsequent anisotropic KOH etching at 80° C to create suspended $SiN_x$ windows under the fine electrodes.

Following cleaning, devices had apertures patterned in the $SiN_x$ via $Ga^+$ focused ion beam (FIB) milling. Single layer graphene was synthesized and transferred by Graphenea at the wafer-scale and a subset of devices were selected for further fabrication. The graphene was then patterned via EBL using negative resist (AR-N 7500) followed by $O_2$ plasma etching to pattern graphene devices between the tips of the electrodes. Devices were then soaked in N-Methyl-2-Pyrrolidone (NMP) followed by acetone and isopropanol (IPA) to remove any residual resist. Finally, devices were pre-screened via scanning electron microscopy (SEM) to select un-damaged devices for subsequent STEM experiments.

**SEEBIC interference correction**
All of the SEEBIC images acquired had persistent interference in them. Several examples are shown in figure S5. We believe this arises from within the microscope column and as such have little recourse to correct it at the acquisition stage. Several techniques were attempted to remove the interference, including Fourier filtering and principal component analysis (PCA), however these techniques either failed to adequately correct for the interference or substantial image information was lost. Eventually, it was discovered that for any image acquired with a given pixel dwell time, the interference pattern was (overwhelmingly) the same. We believe this is the case for two reasons. First, image acquisition is synchronized with the AC mains of the microscope. This is a common feature on electron microscopes that is used to keep interference and noise, as we observed, in sync on all scan lines of the image. Briefly, using this feature ensures that each line waits a very short time before starting, such that it is synchronized with the microscope's AC mains. It is for this reason that the interference takes on a vertical nature in the images. Second, the interference is the same for images of a given dwell time, because the dwell time effectively defines the time it takes to scan across a single line (or row) of the image, thus this defines our sampling of the interference. Since the interference is believed to arise from the microscope itself, the same sampling time in the same microscope would be expected to yield the same interference pattern.

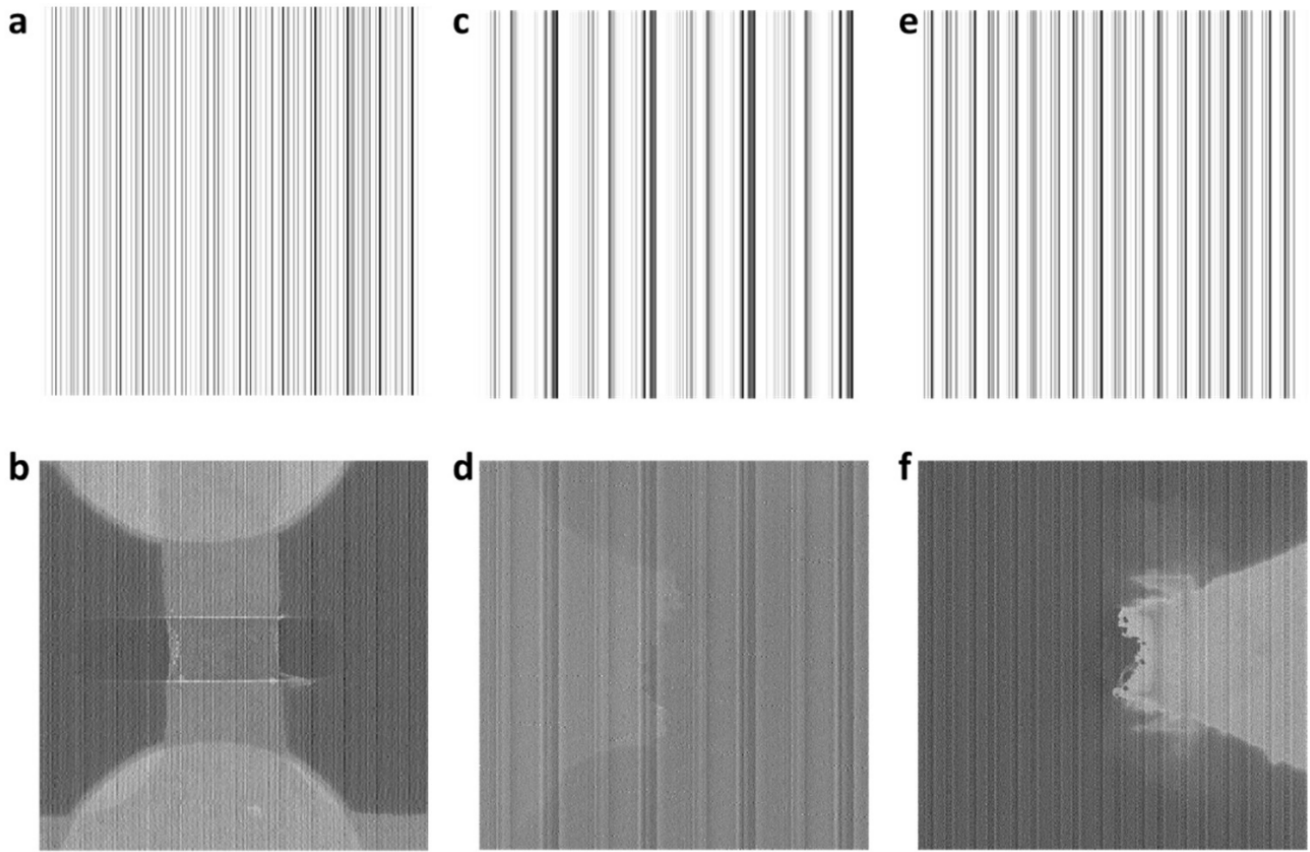

Figure S5: Examples of extracted interference in images for different dwell times (top row) and SEEBIC images with the interference uncorrected (bottom row). Images are pairs (a,b), (c,d), (e,f).

The interference is difficult to correct through conventional filtering methods since it is not just additive but modulates the entire image and is a stronger signal than the SEEBIC signal. Progress was subsequently made on a method to correct the interference when the intensity value was extracted for two separate images of the same dwell time and plotted together, showing that it was effectively the same, see Figure S6.

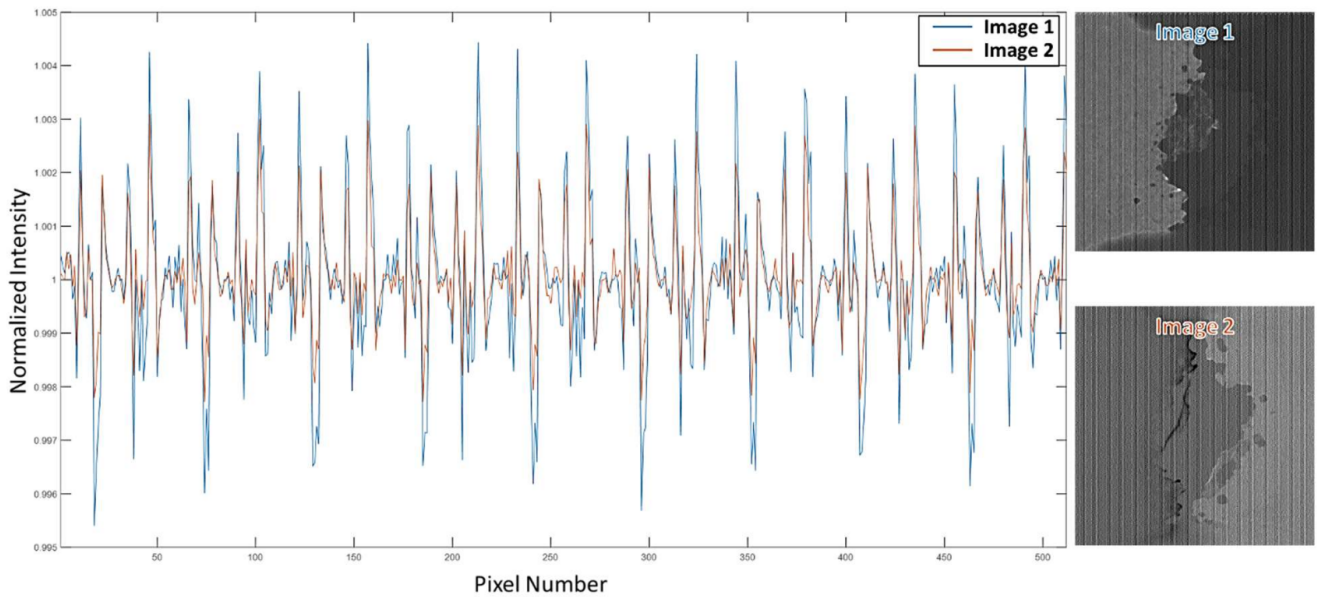

Figure S6: Row interference intensity extracted from two separate example images of graphene devices, shown right, showing substantial alignment between the interference in both images, despite being acquired on separate devices on separate days. Difference in scaling are due to differences in intensity for the images and could be corrected with a scaling factor.

From this knowledge it was possible to write a correction script that was demonstrated to be successful for all images, with a variety of dwell times. The general aspects are shown in Figure S7. Briefly, an image is averaged over all rows to create a line intensity value representing all the pixel intensities in their respective columns. This is shown in Figure S7(b), which already highlights the interference across the image horizontally. From this we can clearly see that it both increases and decreases values. However, this average also captures other aspects of the image, such as intensity variation across the image from the features of interest. To correct for this a background flattening technique was employed to flatten this column averaged line trace. To do this, a 1D moving median filter was used to extract the general shape of the intensity values. It was determined that a moving median filter that was fifty pixels wide best corrected for the interference. This value was not arbitrary and is related to the nature of the interference in the images. Once this general shape was determined, it was used to divide the column averaged line trace to give a flattened version, shown in Figure S7(c). This is the representation of the interference to be corrected. Next, this was reconstructed into an image, by simply using the trace intensity values for all rows in an image of equal dimensions with the image to be corrected. Doing so shows a nice representation of the interference pattern, as can be seen in Figure S7(d). Finally, to correct the images, this interference representation is used to divide the original image, such that regions of high intensity are decreased, and regions of low intensity are increased. This is analogous to gain using a CCD gain reference image to correct images taken on CCDs. This procedure leads to the corrected image shown in Figure S7(f) and has the advantage of not requiring any additional reference images to be acquired.

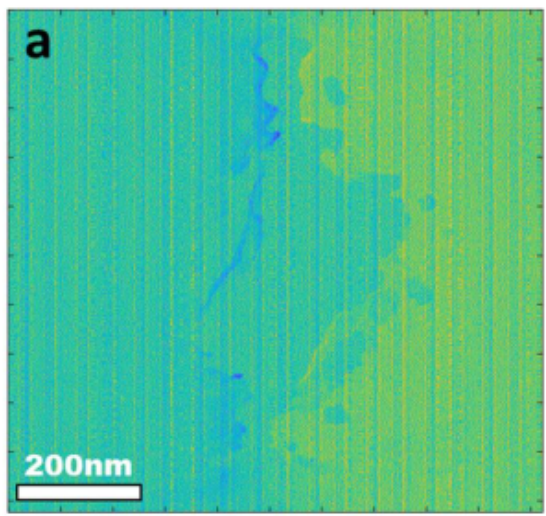
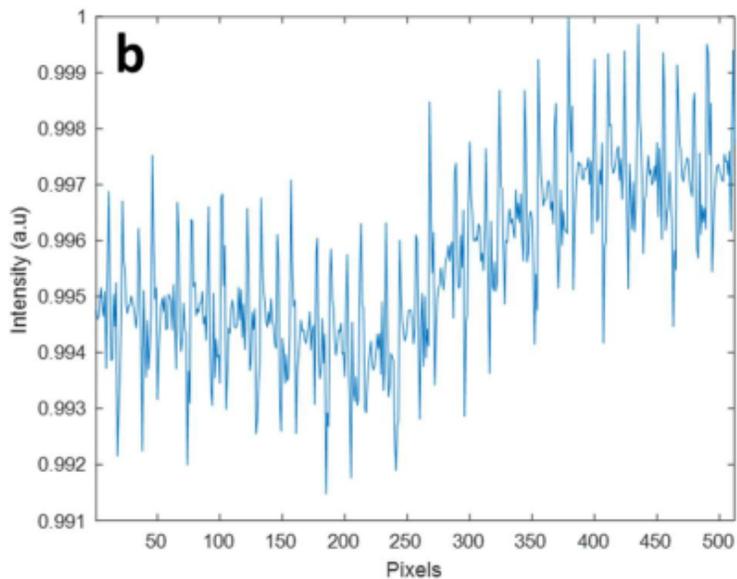
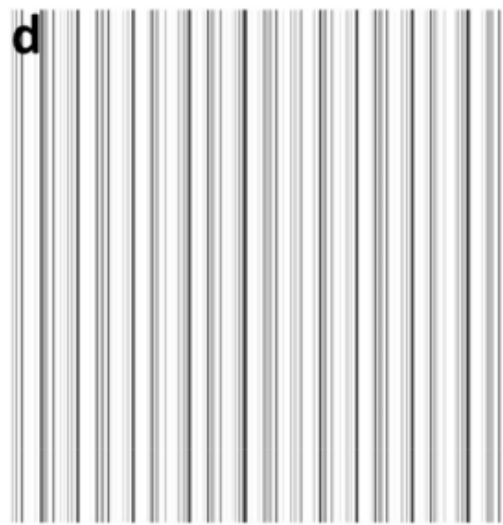
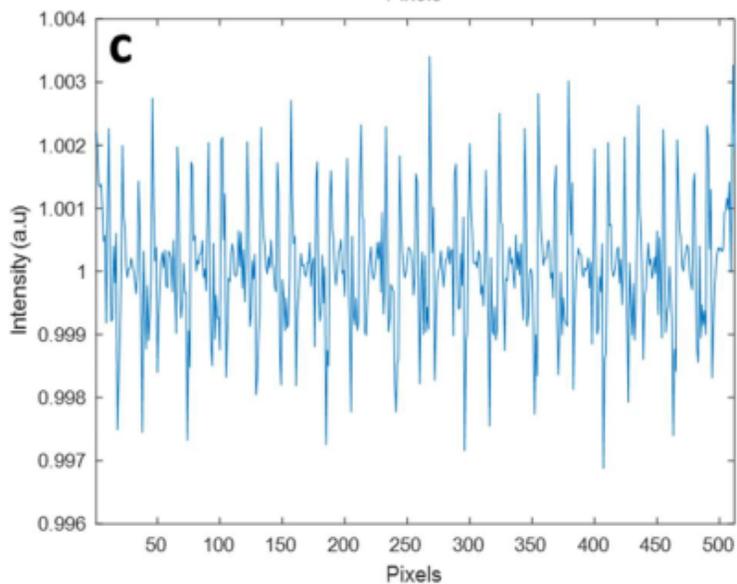
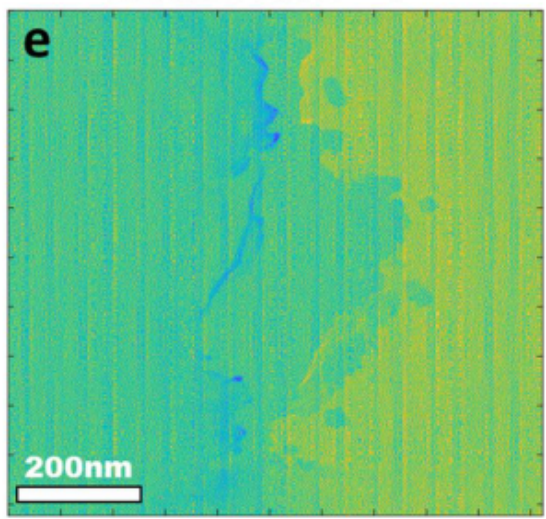
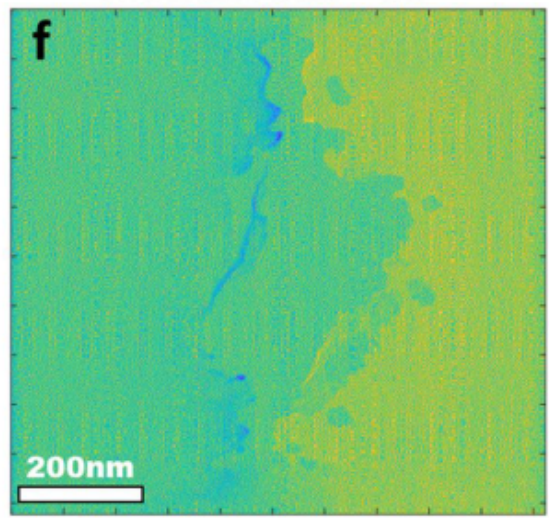

Figure S7: Images were corrected using a background flattening technique with a 'flat' image reconstructed from the extracted image interference. (a) Input image with clear interference running vertically in the image, (b) intensity of all rows averaged together. (c) flattened intensity across the image by using a 1D median filter to extract the general shape to flatten the average, (d) 'flat' image

reconstructed from (c) by extending the line intensity to all rows to create an image, (e) flattened image with (c) using a five pixel wide median filter, (f) flattened image with (c) using a fifty pixel wide median filter.

**Graphene switching behavior extended discussion**

In the main text we highlighted the observation of graphene switching behavior in SEEBIC images as charge flowed through non-ohmic regions of the device. Creation of the non-ohmic region was accomplished using feedback-controlled electric breakdown in the microscope. This strategy enables real-time imaging of the device before, during, and after the electric breakdown process. Conventional imaging (e.g. ADF/BF) allows visualization of some structural changes like recrystalization of the Au electrical contacts and evaporation of the $SiN_x$ support membrane. This gives some insight into the device breakdown process at the high current density limit. However, certain aspects of the device structure are not visible to standard imaging modes. In figure 3 of the main text we show that SEEBIC imaging can reveal discontinuities in the device conductivity. This moves into the field of direct imaging of electrical properties. Using the e-beam we can emit secondary electrons from different spatial locations on the sample and observe the charge flow with the transimpedance amplifier (TIA). On the ohmic devices we observe roughly what amounts to a conductance map. On the non-ohmic devices, however, the removal of charge results in a switching behavior as the current must overcome a conduction barrier before it can flow. To probe this behavior using SEEBIC, we rastered the beam in different directions to differentially remove charge while recording the SEEBIC signal as shown in Figure S8. The scan direction controls the spatial distribution of charge and, because the TIA is connected to only one side of the device, this influences the charge flow through the non-ohmic region.

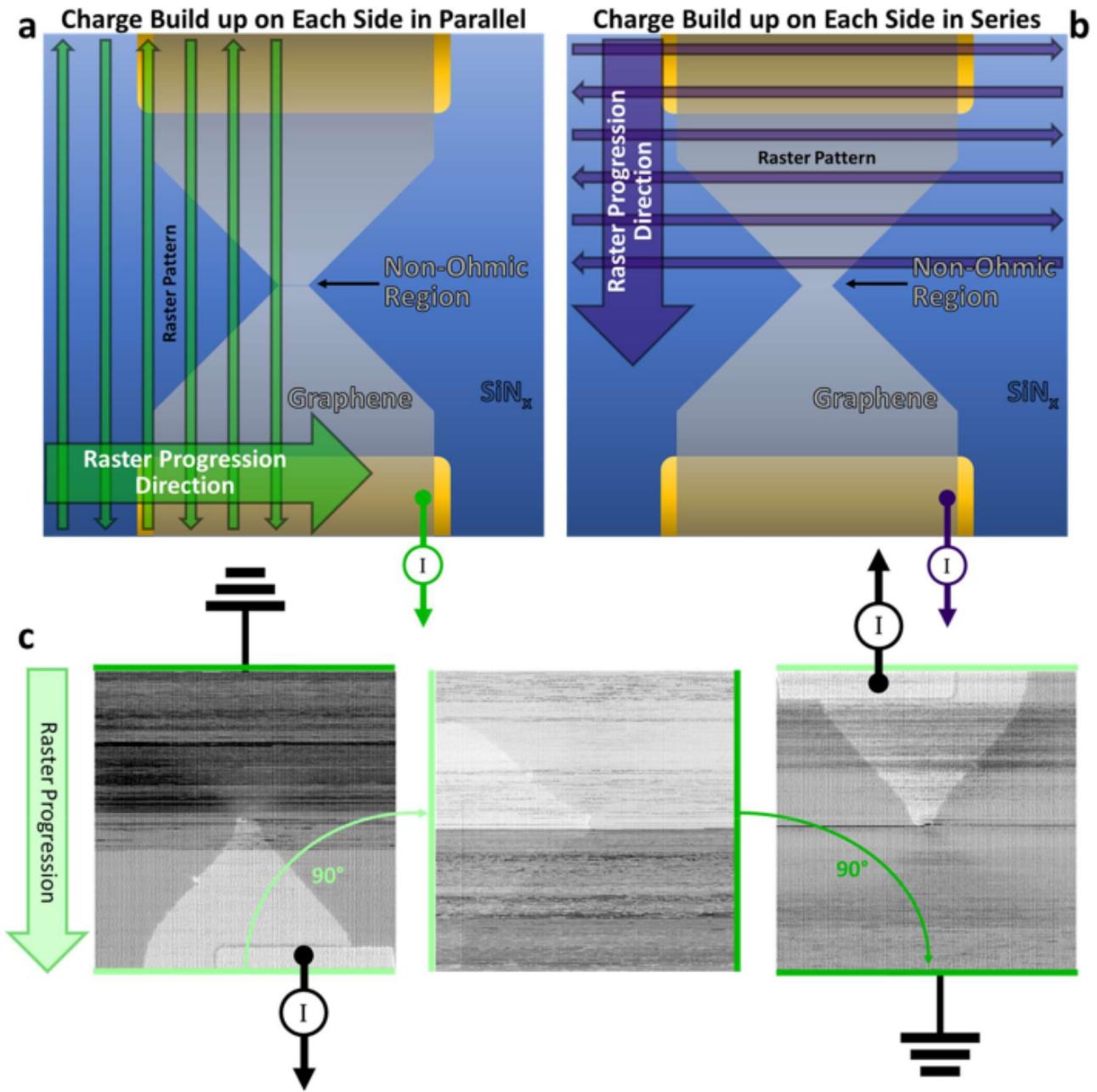

Figure S8: (a),(b) Two examples of possible scan patterns for devices to elucidate aspects of their transport, spatially correlated to ADF imaging. By scanning in different directions, charge and hence voltage and can either be applied to one side of the region of interest or to both sides near simultaneously. (c) examples of the same device with the same connections scanned in three different directions 90° different.

Figure S9 shows how these techniques can be applied to localize the position of a narrow region formed in a graphene constriction through feedback-controlled electric breakdown. The device was initially ohmic as shown in the inset I-V curve in (a), with the narrowest region suspended over an aperture in the SiN$_x$ membrane. After electric breakdown, only a small piece of the ribbon remains connected, shown in the HAADF inset in (b). The I-V curve is also shown inset in (b) showing non-ohmic characteristics. By acquiring two orthogonal SEEBIC images, (c) and (e) (as simple as imaging twice, but changing the scan rotation in between), and overlaying these with the HAADF image, (d), we can observe a dramatic change in the characteristics of the conductance, such that the high resistance region can be inferred, (b). In this instance, the existence of a suspended region permits verification of the narrowest point via HAADF, however this is not possible on all devices since many devices do not permit an aperture beneath them.

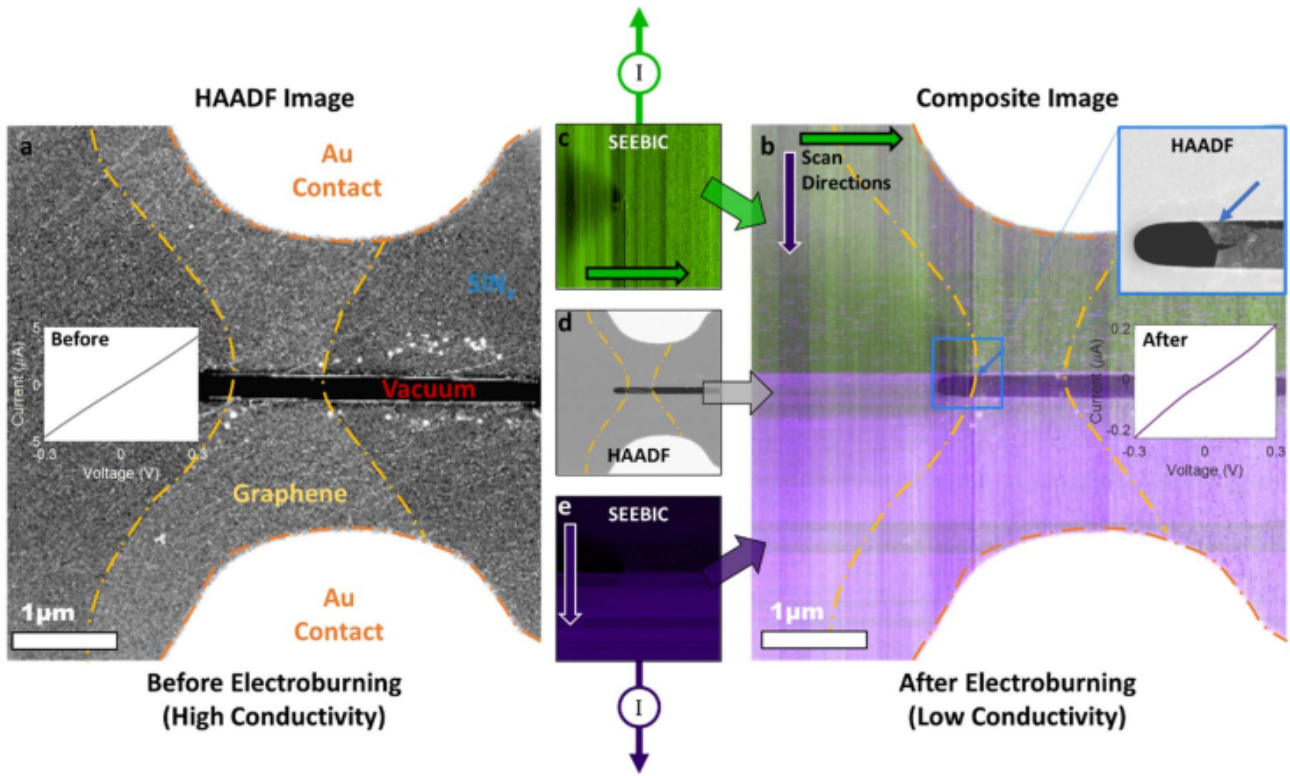

Figure S9: Images acquired at 200kV and 65pA beam current and a 2.5ms dwell time. (a) HAADF image of graphene device (outlined in yellow) connected to two Au contacts and suspended over a SiN x substrate aperture. Inset is an I-V trace of the device prior to electric breakdown to create a high resistivity region. (b) Composite SEEBIC/HAADF image after electric breakdown to create a high resistivity region, comprised of (c) SEEBIC image scanned from left to right, (d) HAADF image, (e) SEEBIC image scanned from top to bottom. Image inset shows the narrow, high resistivity region where switching behavior in the SEEBIC images abruptly changes. I-V inset shows high resistivity after electric breakdown.

The same technique can also be applied to supported devices, which have also undergone electric breakdown to produce high-resistance regions. In the case of the device in Figure S10, electric breakdown resulted in a high-resistivity non-ohmic device, which makes the observed effects more prominent. As in the last device, this device was also SEEBIC imaged twice with the two images having orthogonal dominant scan directions (with simultaneous HAADF images acquired for both). In this instance, the overlay of the two SEEBIC and HAADF images produces an image with four distinct regions. At the center of the quadrants is the highly resistive conduction path between the two graphene halves. For this device, the raw image intensity, representing the SEEBIC signal, has also been included to show how the conduction changes.

The two scanning directions produce distinct SEEBIC signal patterns from the beam induced charging on the two leads of the non-ohmic region and the charging of the substrate. When the beam scans across each lead with every raster, we observed a change in the SEEBIC signal that largely amounts to a step height change in conduction. When the beam scans across each lead in sequence, we observed a high degree of switching behavior when the beam is on the floating lead and then a decrease in switching and conductance when the beam is on the grounded lead.

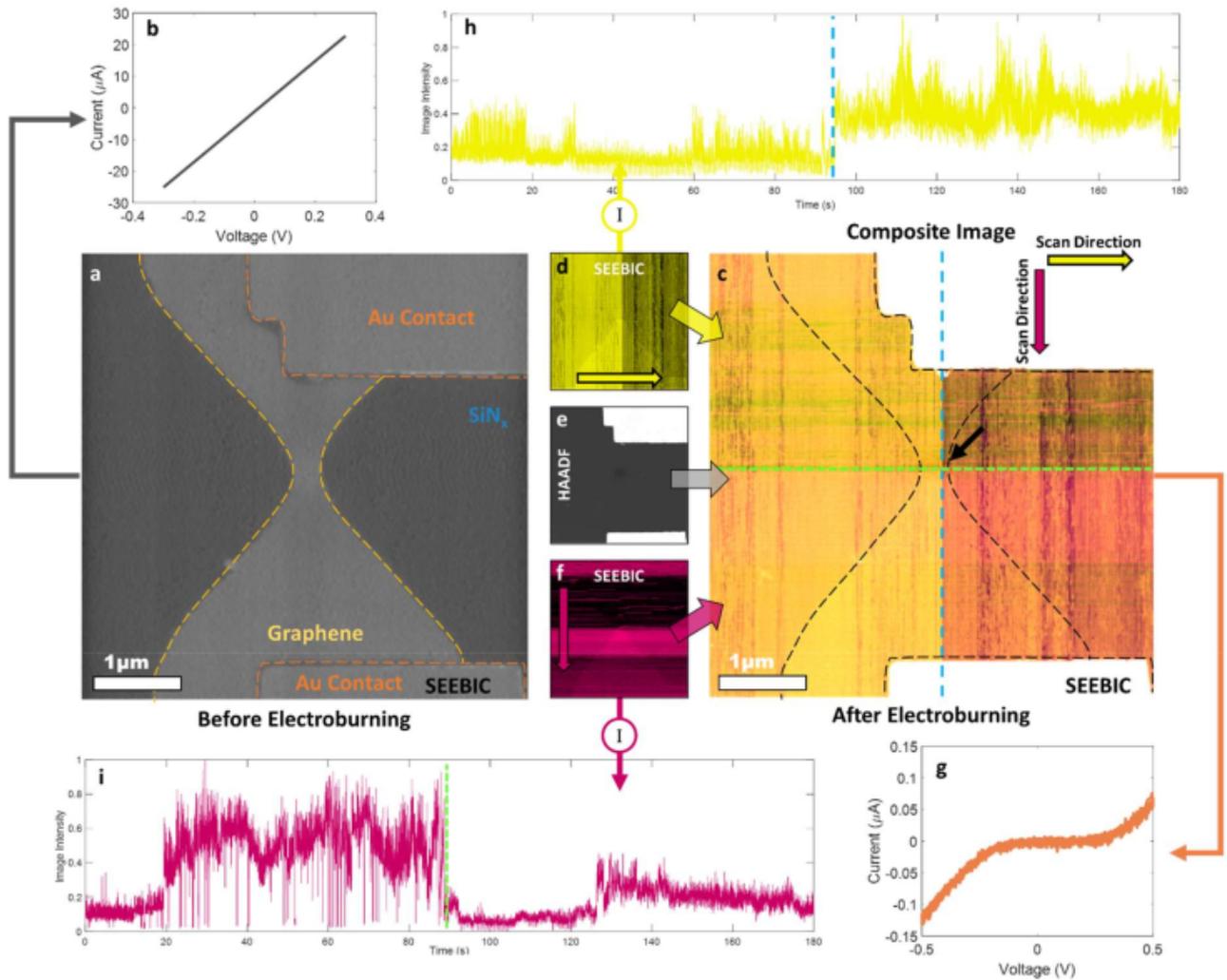

Figure S10: Images acquired at 60kV and 62pA beam current and 2ms dwell time. a) SEEBIC image of graphene constriction (outlined in yellow) connected to Au electrodes (top and bottom, outlined in orange). b) I-V trace of the device prior to electric breakdown, showing high conductivity ohmic behavior across the device. c) Composite SEEBIC/HAADF image comprised of d) SEEBIC image scanned from left to right, e) HAADF image, f) SEEBIC image scanned from top to bottom. The composite of the three images forms an image with four quadrants with the high resistivity region at the center of all four (indicated by black arrow). g) I-V trace of the device after electric breakdown to form a non-ohmic, high resistivity region in the device. h) unraveled I-T trace of SEEBIC image d) showing the transition point (dashed blue line) as the beam scans across the high resistivity region. I) unraveled I-T trace of SEEBIC image f) again showing the transition point (dashed green line) after the beam crosses over the high- resistivity region.

To further explore these phenomena, we applied additional electric breakdown cycles to the device above, to further increase the resistance. The results are shown in Figure S11. Likely due to the substantially higher resistance, this device showed little beam induced current, however a difference in switching was observed as the beam scan direction traversed the high resistivity region. However, in one instance, possibly due to the scan direction and TIA contact configuration, a large current spike was observed as the beam scanned over the region. While possibly a unique result, if repeatable, then this technique could also be used to probe other devices such as quantum dots and single electron transistors.

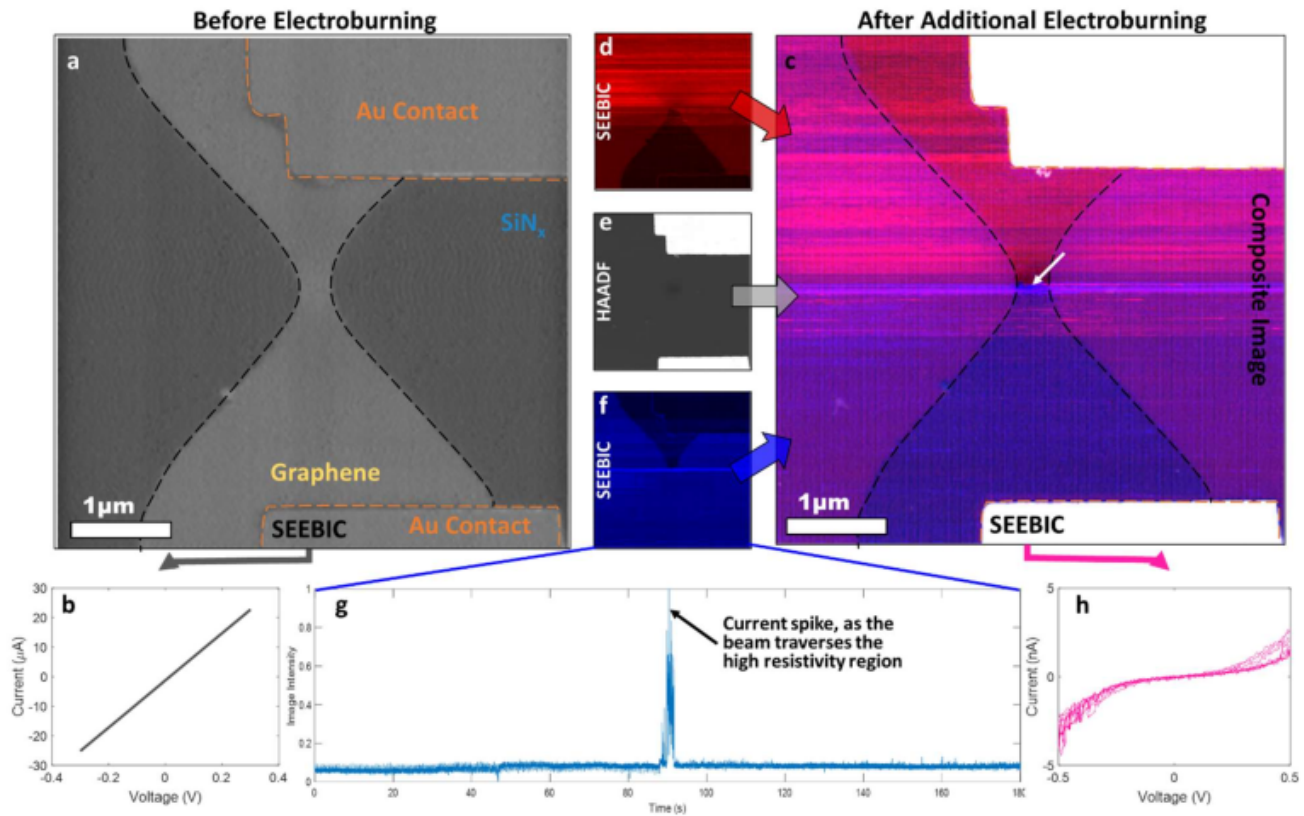

Figure S11: Images acquired at 60kV and 59pA beam current and 2ms dwell time. a) SEEBIC image of graphene constriction prior to electric breakdown. b) I-V trace showing ohmic conduction and high resistivity, c) composite SEEBIC/HAADF image formed from d) SEEBIC image scanned top to bottom, e) HAADF image, f) SEEBIC image scanned from bottom to top. g) I-T trace forming SEEBIC image f), where there is a notable spike where the beam traverses the high resistivity region. h) I-V trace showing non-ohmic transport through the high resistivity region.

In addition to beam induced voltages, one may apply a bias. This has advantages and some considerations. First, to observe the direct beam induced phenomena one must consider the scale of the SEEBIC signal. The bias induced current should be comparable in scale to the anticipated beam induced current, so as to not make the contribution of the SEEBIC signal undetectable. Of course, depending on the beam induced phenomena, the magnitude of the effect could be significant, such as when modulating transport in a non-ohmic device. Additionally, the polarity of the voltage matters. Although many configurations can work, applying the bias such that the two currents are cumulative, increases the overall signal, which optimizes the SNR. The polarity also matters for SE yield, while small voltages in the 100s of millivolts likely have a small impact, applied positive biases in the volt range can suppress SE emission. The advantage of probing a device in situ with an externally applied bias is that it is more representative of typical operation conditions. It is also possible to intentionally apply a large enough bias, so as to diminish the contribution of the SEEBIC current, which would permit the ability to more carefully study substrate gating contributions, which is important for disentangling the various effects.

To demonstrate the feasibility of such an approach, the device from Figure S11 was imaged again using a modified SEEBIC configuration with a +100mV bias applied on the electrode opposite the TIA, shown in Figure S12. Again, a SEEBIC image is produced, but the conductive graphene can no longer be observed, instead only a global change in the current is detected as the beam crosses the resistive region. This suggests that, consistent with previously presented observations, many of the SEs contributing to the current arise from the substrate. The proportion of the SEEBIC image intensity also gives an approximate sense for the beam induced voltages present, relative to the applied +100mV bias.

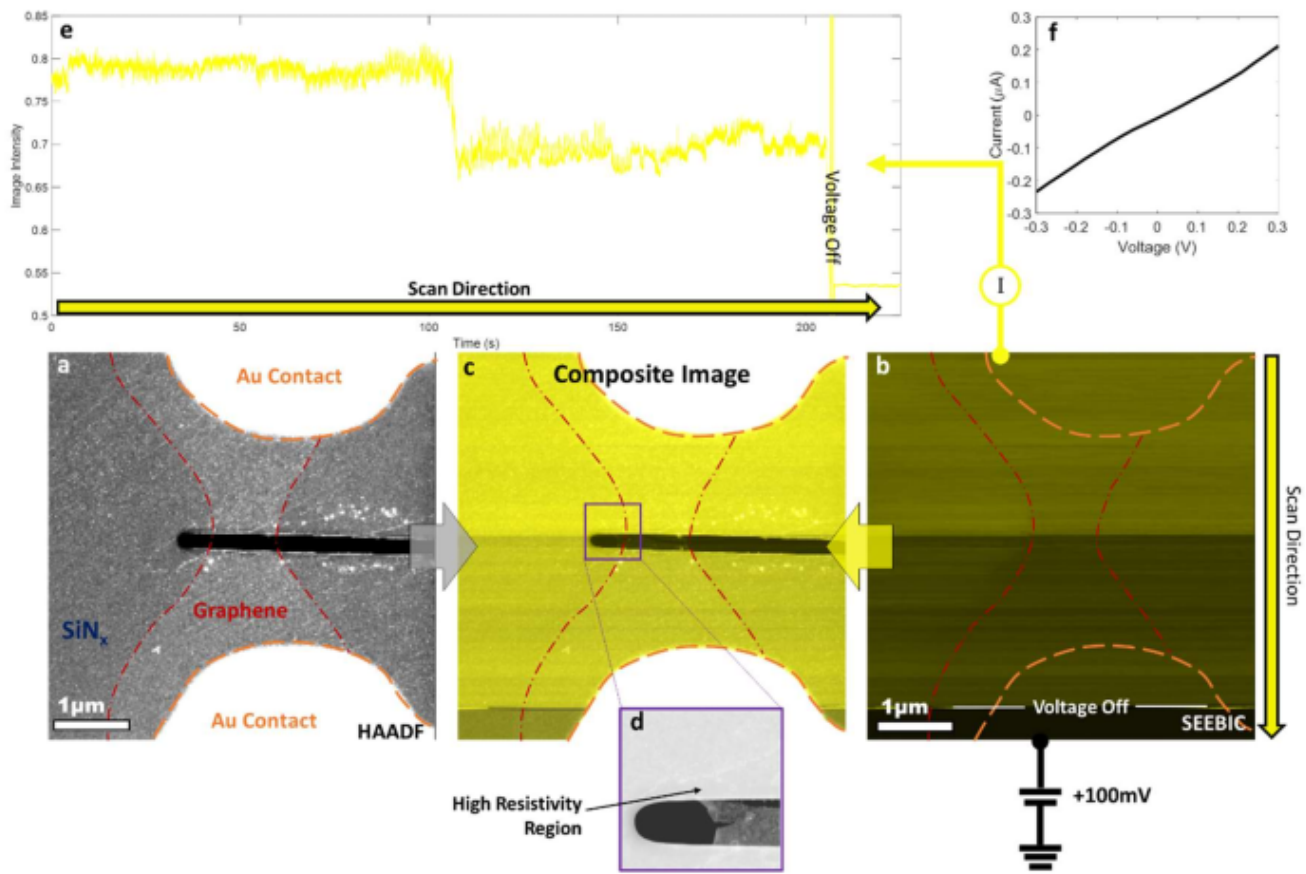

Figure S12: Biased resistive contrast imaging. Images acquired at 200kV and 65pA beam current and 2.5ms dwell time. a) HAADF image of partially suspended graphene constriction (outlined in red) after electric breakdown to create a high resistivity region, connected to two gold contacts on the top and bottom (outlined in orange), b) SEEBIC image taken with +100mV bias on the bottom contact, where a clear transition is shown partially through the image, as wells as towards the bottom, which correlates to the bias voltage being set to ground again. c) composite image of a) and b) showing the correlation between the high resistivity region and the change in intensity in the SEEBIC image. d) higher magnification HAADF image showing the point where the graphene is narrowest corresponding to the change in SEEBIC signal intensity. e) raw SEEBIC signal intensity acquired during imaging, showing a step change in current after the high resistivity region and another step change when the bias voltage was turned off, f) I-V trace showing resistivity of the device.

In Figure 4(e) of the main text we showed a pair of SEEBIC images exhibiting switching behavior with the TIA connected to either side of the device. In Figure 4(f) we showed a plot of the SEEBIC intensity through time over a portion of the upper image. In this profile we observe conductance switching primarily between two states. While this illustrates reversible conductance switching over a long period of time, many other examples switch between more than two states, often more quickly. In Figure S13 we show profiles taken from the complementary image (Figure 4(e) bottom). In particular, in each profile we show just a few image scan lines so that the datapoints are more highly dispersed than the plot shown in Figure 4(f). To the left of the SEEBIC image, we show a plot of each pixel intensity (horizontal axis) vs acquisition time (vertical axis) scaled so that it aligns with the image. This gives an overview of the full range of observed intensities. On the right, is displayed profiles from various regions of the image, composed of only 3-6 scan lines, represented by the different colors.

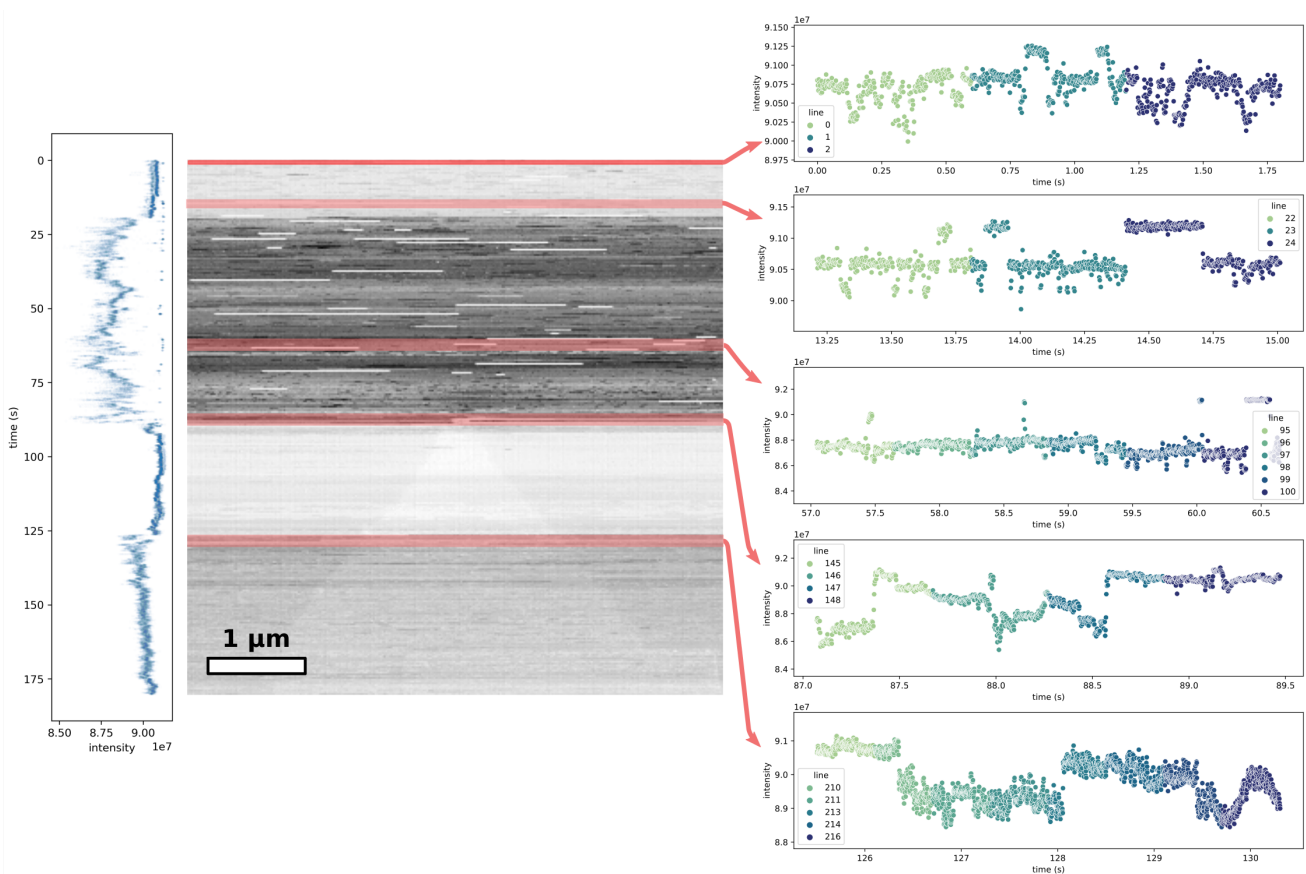

Figure S13 SEEBIC intensity profiles observed during graphene conductance switching behavior. Left: SEEBIC intensity (horizontal axis) vs time (vertical axis). Right: SEEBIC intensity (vertical axis) vs time (horizontal axis) taken from the indicated regions. Color represents the scan line index.